\documentclass[epj]{webofc}
\usepackage[varg]{txfonts}   

\woctitle{MESON2016 - the 14$^\textrm{th}$ International Workshop on Meson Production, Properties and Interaction}
\begin{document}
\selectlanguage{english}
\title{Theoretical studies of the reaction $e^+ e^- \to K^+ K^- \gamma$}

\author{Leonard Le\'sniak\inst{1}\fnsep\thanks{\email{Leonard.Lesniak@ifj.edu.pl}} \and
        Micha\l{} Silarski\inst{2}
}

\institute{Institute of Physics, Jagiellonian University, 30-348 Krak\'ow, Poland 
\and
           Laboratori Nazionali di Frascati dell'INFN, Frascati, Italy
          }

\abstract{ 
We present studies of the $e^+ e^- \to K^+ K^- \gamma$ reaction for the $e^+e^-$ center-of-mass energies close to the $\phi(1020)$ meson mass. Different mechanisms leading to the final state are considered.
The strong interaction amplitudes of the $K^+K^-$ pairs in the $S$-wave are taken into 
account. In addition, the photon emission in the initial state, the final state radiation effects and
all the possible interference terms are included in the transition 
matrix elements. The $K^+K^-$  effective mass distributions and the angular
dependence of the reaction cross-section are calculated. 
}
\maketitle
\section{Introduction}
\label{intro}
A measurement of the $e^+e^- \rightarrow K^+ K^-\gamma$ transition could provide a new information about the $K\bar{K}$ strong interactions near threshold as well as about the structure of the $f_0$(980) and $a_0$(980) scalar mesons. 
The kaon-antikaon interaction was studied in several proceses, e.g. in the $pp \to pp K^+ K^-$ reaction
near the kinematical threshold but the strong interactions of protons in the final state made the interpretation of the results very difficult~\cite{PhysRevC,physRevC2,anke_last}. There are no such obstacles in the $K\bar{K}\gamma$ system. 
For the $e^+e^- \rightarrow K^{0}\bar{K^{0}}\gamma$ branching fraction only the upper limit $1.9~\cdot 10^{-8}$ is known ~\cite{KLOE 2009} and the branching fraction for the $\phi$(1020) meson decay into the $K^+K^-\gamma$ channel is yet unknown.
The $\phi(1020) \rightarrow \pi^+\pi^-\gamma$ branching fraction has been measured: $\Gamma(\pi^+\pi^-\gamma)/\Gamma_{total} = (4.1 \pm 1.3) \cdot 10^{-5}$ ~\cite{CMD2}
 and the fits for the branching fractions of the $\phi$(1020) meson decay into the two scalar resonances plus
photon give the following results: $\Gamma( f_{0}(980)\gamma)/\Gamma_{total} = (3.22 \pm 0.19) \cdot 10^{-4}$, $\Gamma(a_{0}(980)\gamma)/\Gamma_{total} = (7.6 \pm 0.6) \cdot 10^{-5}$~\cite{PDG}.
Both $f_0$ and $a_0$ decay to the kaon pairs with low relative momenta, thus studies of the $e^+e^- \rightarrow K^+ K^-\gamma$ transition at the $\phi$ mass peak could provide a new information about the $K^+K^-$ strong interactions near threshold as well as about the structure of the scalar mesons.

\section{$K^+K^-$ effective mass distributions and angular distributions}
\label{sec-2}
The differential cross section for the reaction
$e^+(p_{e^+}) e^-(p_{e^-}) \to K^+(p_{K^+}) K^-(p_{K^-})\gamma(p_{\gamma})$
reads 
\begin{equation}
\label{dsig}
\frac{d\sigma}{dm^2 dm^2_{K^-{\gamma}}dtdt_1}=\frac{|M|^2}{(2\pi)^4 16 s (s-4 m^2_e) (s-m^2)r}.
\end{equation}
The matrix element $M$ can be expressed in terms of five invariants depending on the particle momenta $p$: $s=(p_{e^+}+p_{e^-})^2$, two effective masses squared
$m^2=(p_{K^+}+p_{K^-})^2$ and $m^2_{K^-\gamma}=(p_{K^-}+p_{\gamma})^2$, and two momentum transfers squared
$t=(p_{e^-}-p_{\gamma})^2$ and $t_1=(p_{e^-}-p_{K^-})^2$. In Eq.~(\ref{dsig})
$r=\sqrt{-(t_1-t_{1min})(t_1-t_{1max})}$, where  $t_{1min}$ and $t_{1max}$ are the kinematically allowed
minimum and maximum values of the momentum transfer squared $t_1$, respectively.
The squares of the momentum transfers $t_1$ and $t$ are related to the $K^-$ and photon polar angles 
$\theta_1$ and $\theta_{\gamma}$, defined in the $e^+ e^-$ center-of-mass frame with the z-axis chosen along the $e^-$ momentum:

\begin{equation}
\label{t1t}
t_1\approx m_K^2-\sqrt{s}~ E_1^l (1-v_1 \cos{\theta_1}),~~~~~~~~~~~~~~~
t\approx -\sqrt{s}~ \omega^l (1-\cos{\theta_{\gamma}}).
\end{equation}
In the above equations, where a small value of the electron mass is neglected, $v_1$ is the $K^-$ velocity, $E_1^l$ and $\omega^l$ are the $K^-$ and photon energies
given by the following expressions:
\begin{equation}
\label{Eomega}
E_1^l=\frac{m^2+m_{K^-\gamma}^2-m_K^2}{2\sqrt{s}},~~~~~~~~~~~~~\omega^l=\frac{s-m^2}{2\sqrt{s}}.
\end{equation}

The $K^-\gamma$ effective mass is related to the $K^-$ polar angle $\theta_1^*$ defined
with respect to the direction of the photon momentum in the $K^+K^-$ center-of-mass frame
($v$ being the $K^-$ velocity):

\begin{equation}
\label{zet}
m_{K^-\gamma}^2=m_K^2+\frac{1}{2}(s-m^2)(1-v~z),~~~~v=\sqrt{1-4m_K^2/m^2}, ~~~z=\cos{\theta_1^*}.
\end{equation}

In Ref.~\cite{pipi} two theoretical models have been exploited in detailed fits to the KLOE data of the 
$\phi \to f_0(980) \gamma \to \pi^+ \pi^- \gamma$ decay. The first model labeled NS (no-structure model) has been formulated by Isidori, Maiani, Nicolaci and Pacetti in Ref.~\cite{NS}. The second one
called the kaon loop model has been developed by Achasov and his collaborators (see for example ~\cite{KL}). In this study we present results 
corresponding to the NS-model of the $e^+ e^- \to K^+ K^- \gamma$ reaction. The parameters of the model have been taken from Table 1 of ~\cite{pipi}.

\begin{figure}[htp]
\includegraphics[scale = 0.28]{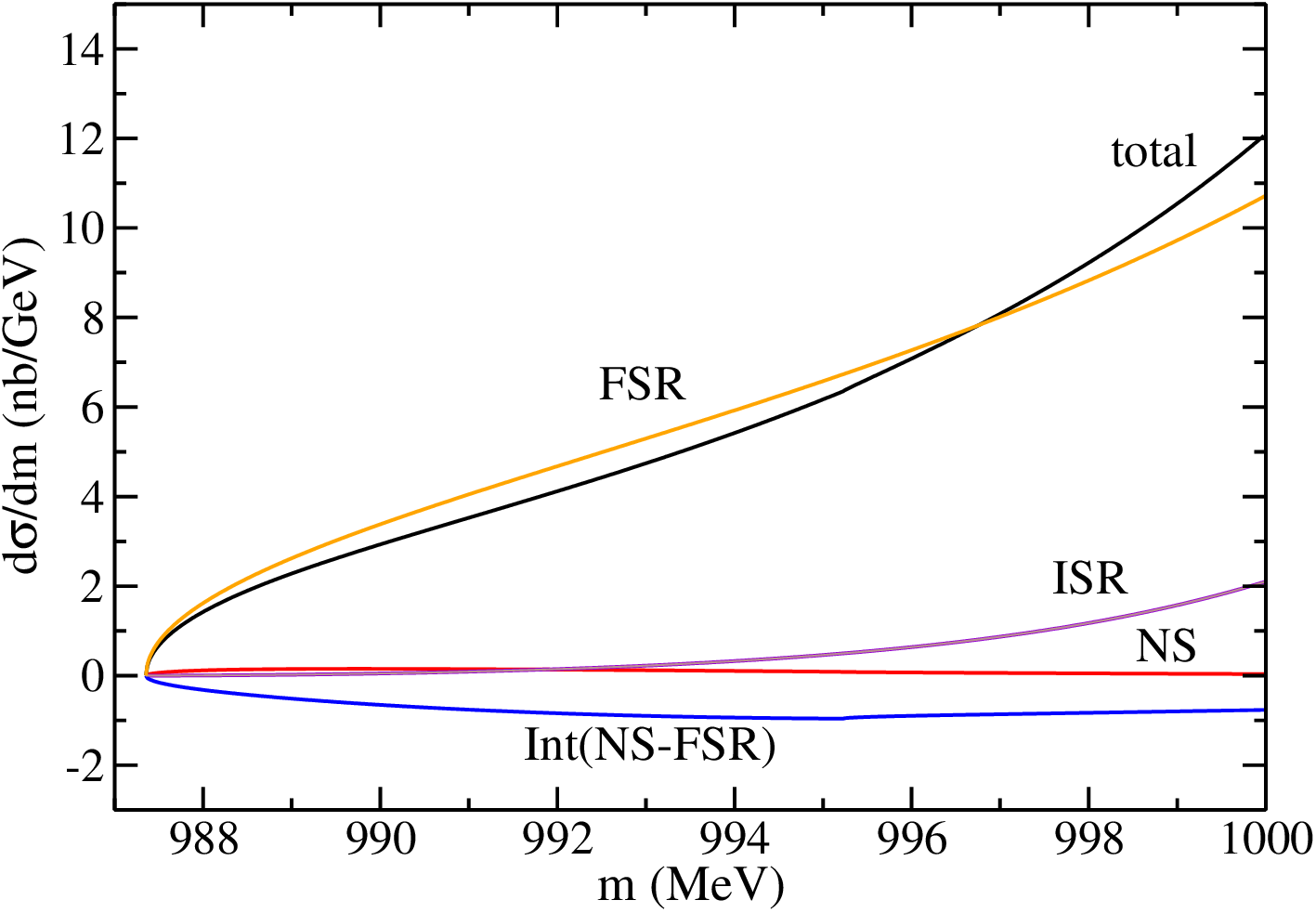}~~~~~
\includegraphics[scale = 0.28]{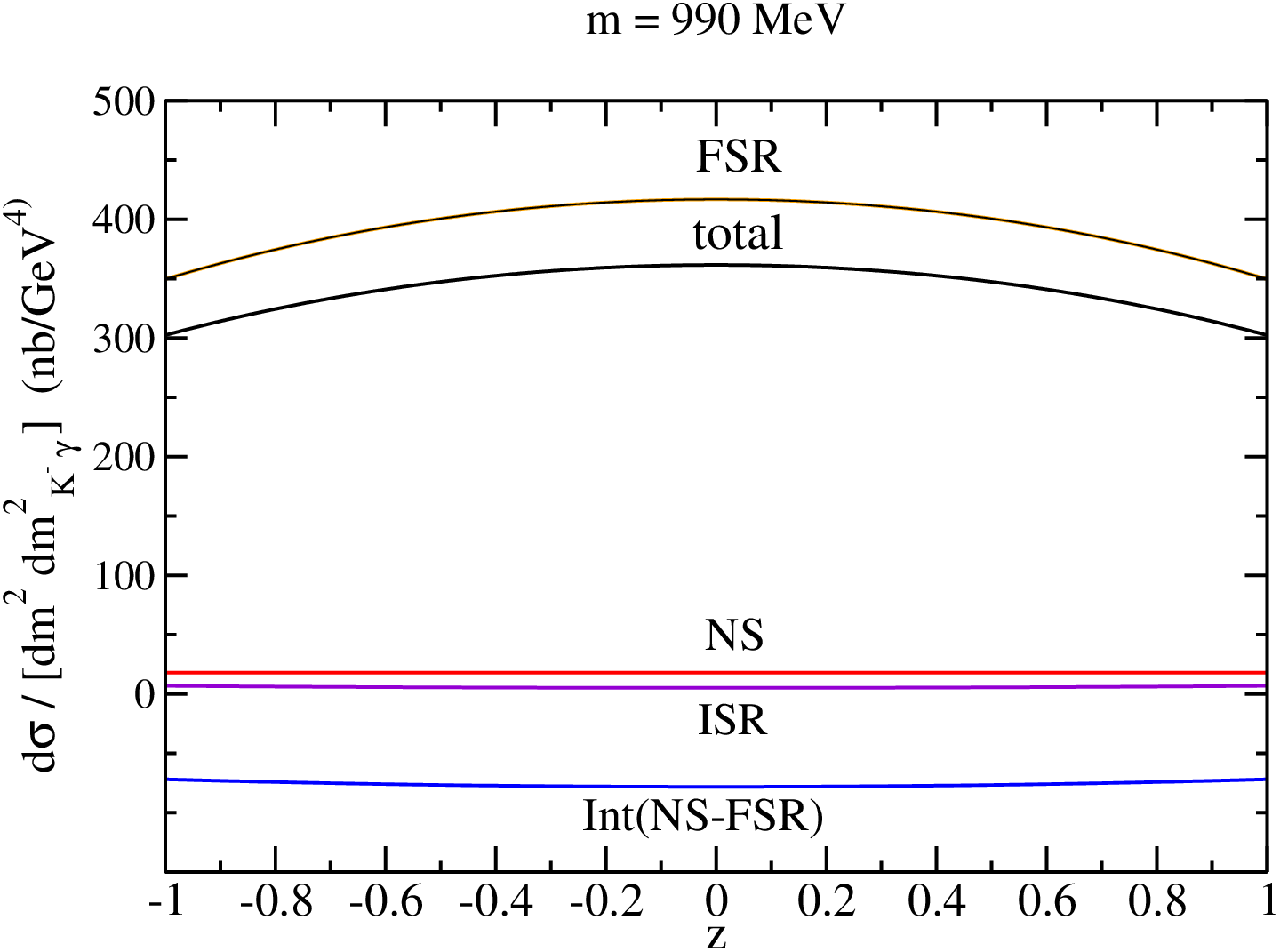}~~~~~
\centering
\caption{Left panel: $K^+K^-$ effective mass distributions,
right panel: double differential cross sections at m=990 MeV versus the cosine of the $K^-$ polar angle defined in Eq.~(\ref{zet}). In both cases 
$45^0 < \theta_{\gamma} < 135^0.$
Four contributions to the total result are presented. They are labeled by FSR (final state radiation), ISR (initial state radiation), NS (NS model) and Int(NS-FSR) (interference of the NS and the FSR amplitudes).
}
\label{fig-1}       
\end{figure}

\begin{figure}[ht]
\centering
\includegraphics[scale = 0.28]{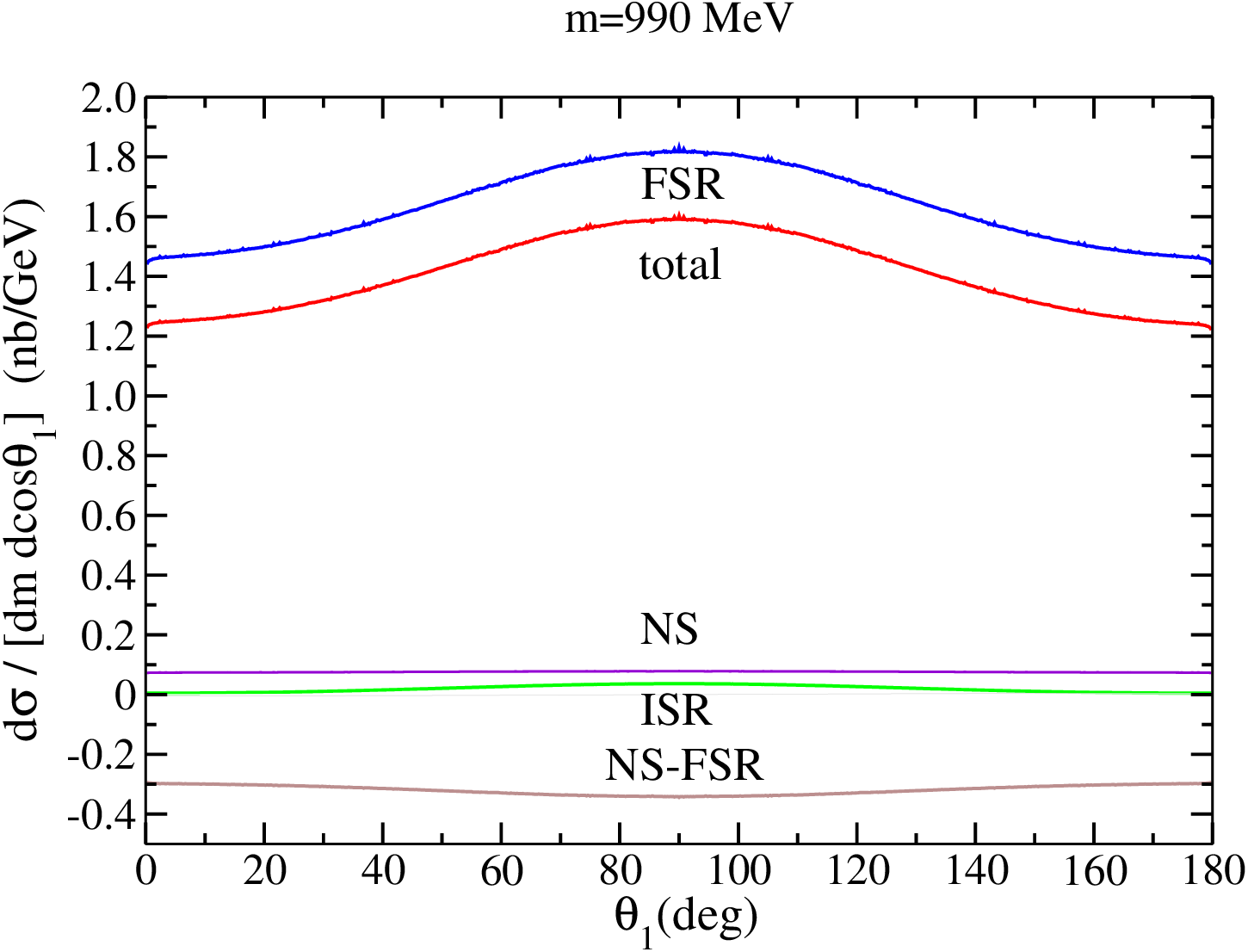}~~~~
\includegraphics[scale = 0.28]{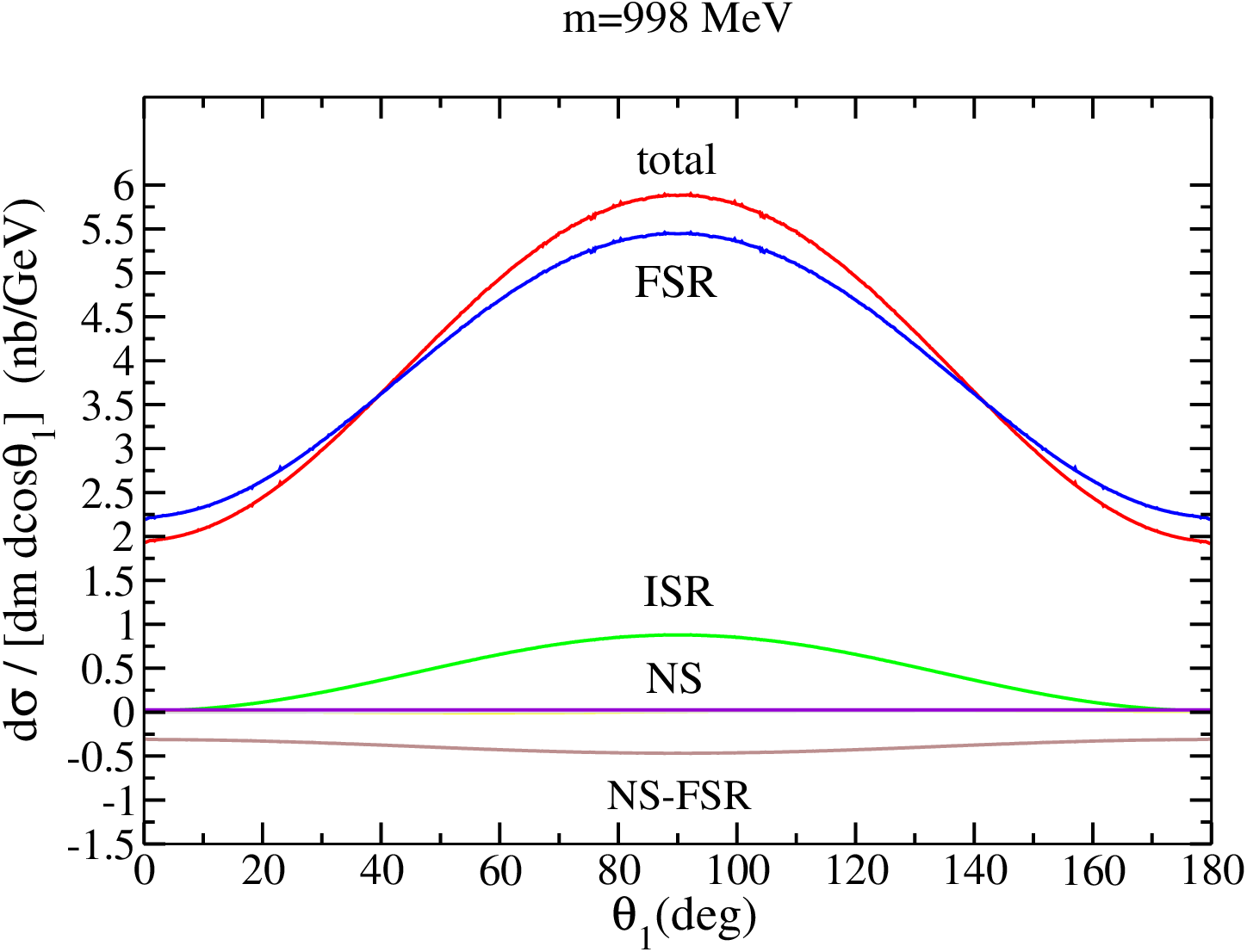}~~~~\\
\includegraphics[scale = 0.275]{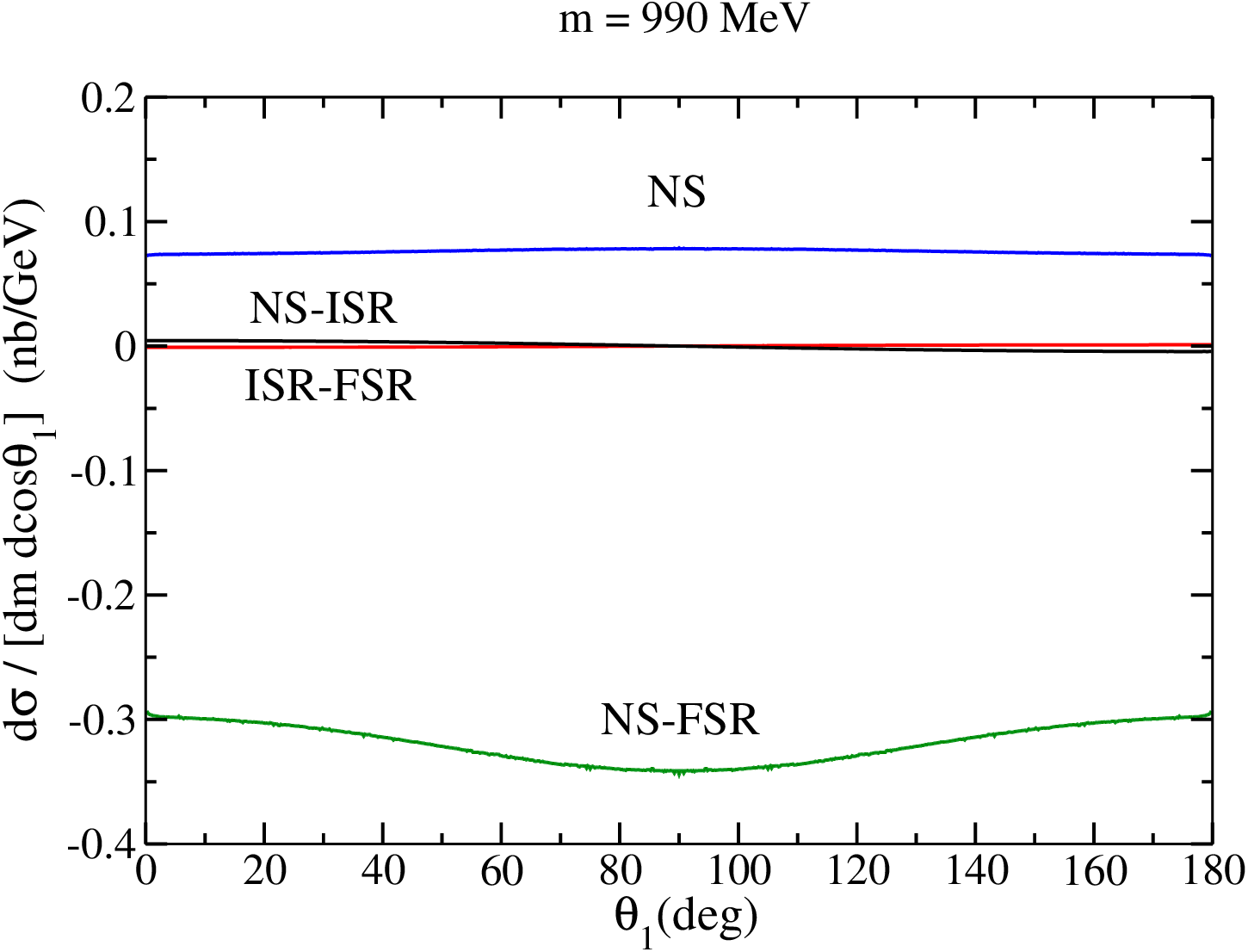}~~~~
\includegraphics[scale = 0.275]{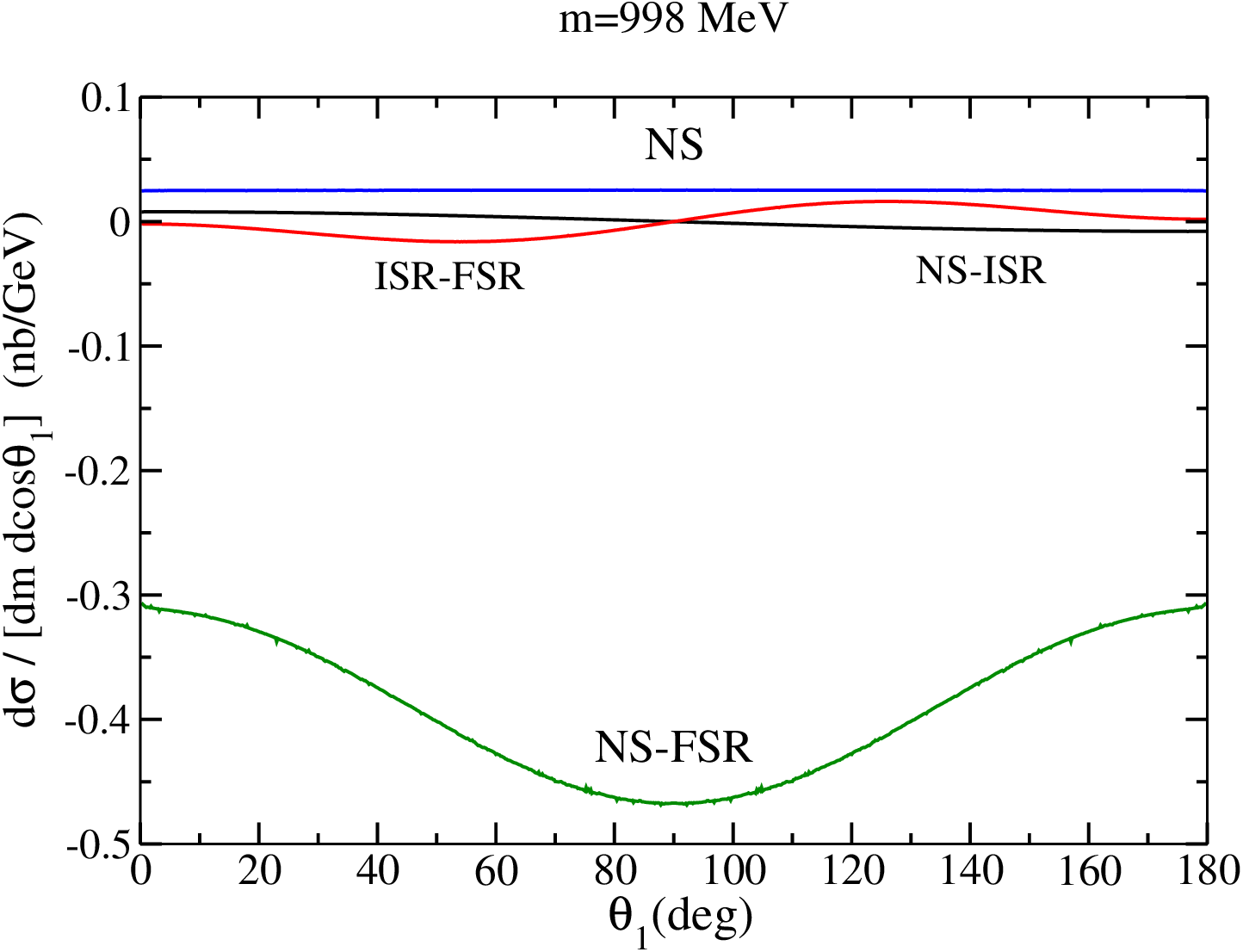}~~~~
\caption{$K^-$ angular distributions at fixed m for $45^0 < \theta_{\gamma} < 135^0$. The meaning of the labels total, FSR, ISR and NS is the same as in Fig.~\ref{fig-1}.
By NS-FSR, ISR-FSR and NS-ISR we denote the interference terms of the NS and FSR amplitudes,
the ISR and FSR amplitudes, and the NS and ISR amplitudes, respectively. 
}
\label{fig-2}      
\end{figure}

\begin{figure}[ht] 
\centering
\includegraphics[scale = 0.28]{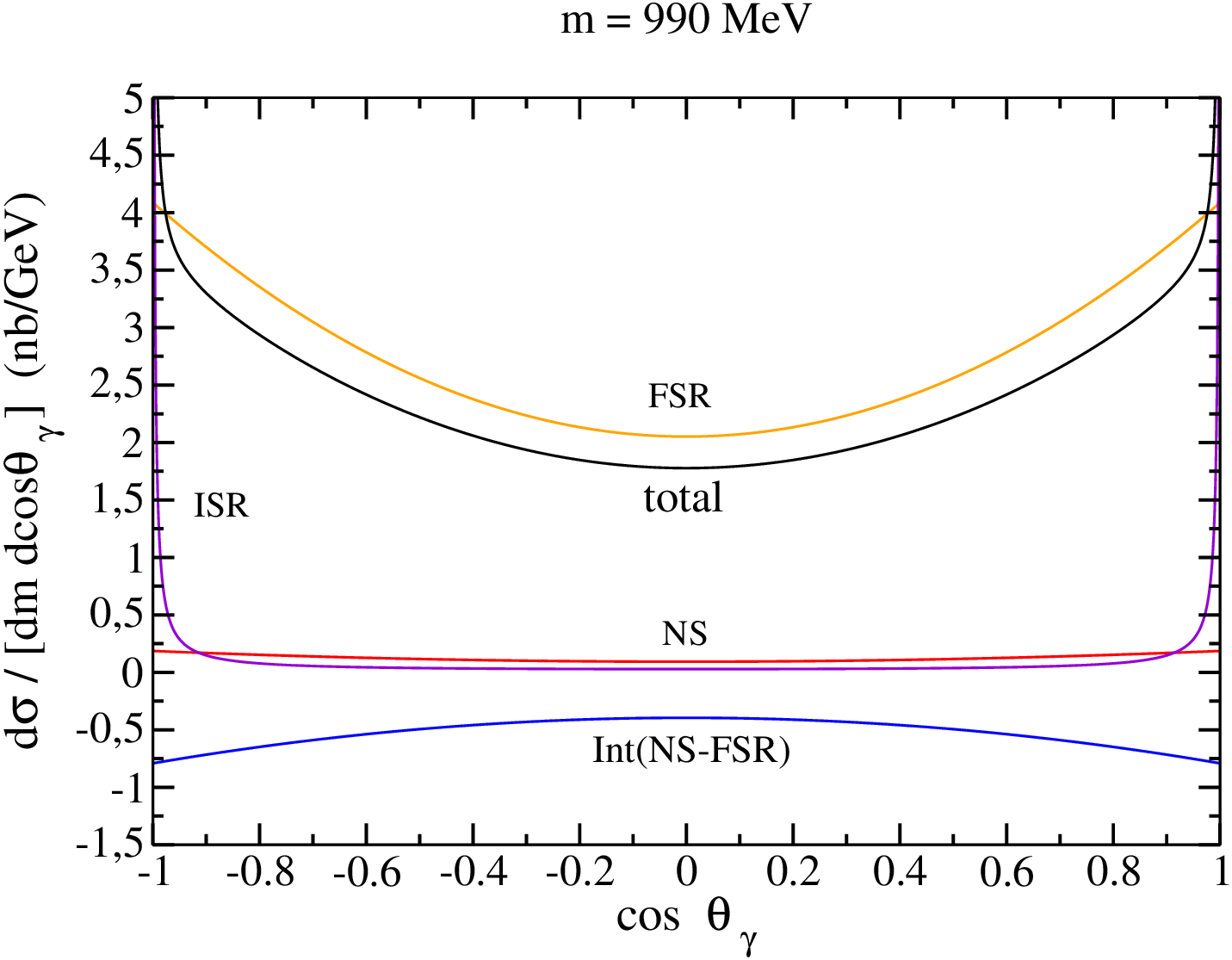}~~~~
\includegraphics[scale = 0.28]{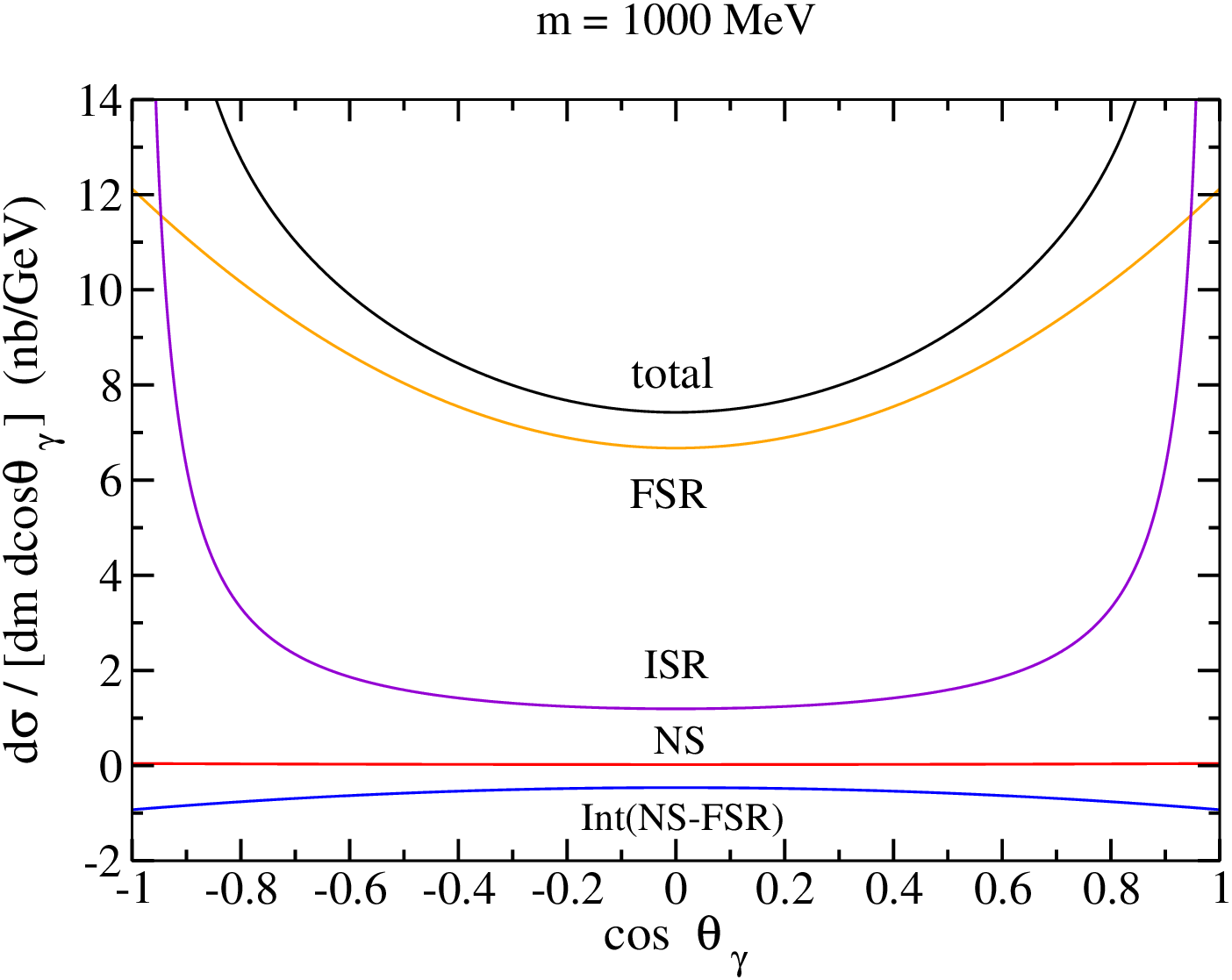}~~~~
\caption{Photon angular distributions at fixed m=990 and 1000 MeV.}
\label{fig-3}      
\end{figure}

In Fig.~\ref{fig-1} the $K^+K^-$ effective mass distributions and $K^-$ angular distributions with respect to the photon momentum axis in the $K^+K^-$ center-of-mass frame are jointly shown.  The $K^-$ angular distributions in the $e^+e^-$ c.m. frame are shown in Fig.~\ref{fig-2} for two $K^+K^-$ effective masses $m$ equal to 900 and 998 MeV. In Fig.~\ref{fig-3} the photon angular distributions are presented 
for two values of $m$.

\begin{table}[h]
\centering
\caption{The cross sections integrated over the $K^+K^-$ effective mass from the threshold up to 1009 MeV 
for two ranges of the photon emission angle $\theta_{\gamma}$. 
}
\label{tab-1}       
\begin{tabular}{lll}
\hline
reaction mechanism & $24^0 < \theta_{\gamma} < 156^0$ & $45^0 < \theta_{\gamma} < 135^0 $\\\hline
FSR                &  0.330 nb                        &    0.238 nb\\
NS                 & 0.0020 nb                        &    0.0014 nb\\
Int(NS-FSR)        & -0.021  nb                        &    -0.015 nb\\
ISR                & 0.183  nb                        &    0.104 nb\\\hline
total              & 0.494  nb                        &    0.328 nb\\\hline
\end{tabular}
\end{table}

\section{Prospects of the $e^+ e^- \to K^+ K^- \gamma$ reaction studies with the KLOE data}
\label{sec-3}
One of the best data sets which could be used to study the $e^+e^- \to K^+ K^-\to K^+ K^-\gamma$ reaction was collected by the KLOE experiment operating at the DA$\Phi$NE $\phi$-factory in Frascati, Italy~\cite{kloe2008}.
 Its advantage is a very good kaon momenta
determination and the high statistics. However, one needs to deal with a detection of low energy photons ($E_{\gamma} < $ 32 MeV in the $e^+ e^-$ c.m. system) and the slow kaon momentum tracks at the low $K^+K^-$ effective masses. Based on the values of the cross sections given in Table~\ref{tab-1} we show in Table ~\ref{tab-2} the estimated numbers of events for the integrated lumonisty of 1.7 fb$^{-1}$. Corrections for the detector acceptance and 
efficiency are 
not yet done. 

\begin{table}[h]
\centering
\caption{Numbers of events  
for the $K^+K^-$ effective mass up to 1009 MeV and for two ranges of $\theta_{\gamma}$.
}
\label{tab-2}       
\begin{tabular}{lll}
\hline
reaction mechanism & $24^0 < \theta_{\gamma} < 156^0$ & $45^0 < \theta_{\gamma} < 135^0 $\\\hline
FSR                & 5.6$\cdot  10^5$                        &   4.0$\cdot  10^5$ \\
NS                 & 3.4$\cdot  10^3$                        &   2.4$\cdot  10^3$ \\
Int(NS-FSR)        & -3.6$\cdot  10^4$                        &   -2.5$\cdot  10^4$ \\
ISR                & 3.1$\cdot  10^5$                        &   1.8$\cdot  10^5$ \\\hline
total              & 8.4$\cdot  10^5$                        &   5.6$\cdot  10^5$ \\\hline
\end{tabular}
\end{table}

These results of theoretical calculations can be used in a future experimental measurement 
of the unknown branching fraction of the $\phi(1020)$ meson
into the $K^+K^-\gamma$ channel.

\begin{acknowledgement}

This work has been supported by the Polish National Science Centre (grant no 2013/11/B/ST2/04245).
\end{acknowledgement}

\end{document}